# Theoretical determination of the Raman spectra of MgSiO$_3$ perovskite and post-perovskite at high pressure

Razvan Caracas and Ronald E Cohen

Geophysical Laboratory, Carnegie Institution of Washington,

Washington DC 20015, USA, E-mail: r.caracas@gl.ciw.edu

**Abstract**

We use the density functional perturbation theory to determine for the first time the pressure evolution of the Raman intensities for a mineral, the two high-pressure structures of MgSiO$_3$ perovskite and post-perovskite. At high pressures, the Raman powder spectra reveals three main peaks for the perovskite structure and one main peak for the post-perovskite structure. Due to the large differences in the spectra of the two phases Raman spectroscopy can be used as a good experimental indication of the phase transition.

Introduction

The Earth's mantle, a complex mixture of silicates and oxides, accounts for about half of the total Earth mass. The mineralogical composition of the mantle simplifies with increasing pressure and temperature, and the different discontinuities identified by seismic analysis usually correspond to mineral phase transformations. In the lower mantle, the region in the Earth extending from about 660 km to about 2900 km depth, it is generally believed that the main mineral phase is Fe-bearing magnesium silicate, which can make up to about 70 volume %.

It was long thought that $(Mg,Fe)SiO_3$ is stable down to the base of the mantle (corresponding to pressures on the order of 135 GPa) in the perovskite structure. Recently, a new phase transition in $MgSiO_3$ perovskite has been discovered experimentally [*Murakami et al.*, 2004, *Oganov and Ono*, 2004] and confirmed theoretically [*Oganov and Ono,* 2004, *Tsuchiya et al.*, 2004a]. The transition occurs at around 125 GPa and 2500 K and is characterized by a Clapeyron slope of 7 kbar/K [*Tsuchiya et al.*, 2004a]. The new high-pressure phase, post-perovskite, has Cmcm symmetry and is isomorphous to $CaIrO_3$. The newly discovered phase transition explains most of the seismic anomalies [*Oganov and Ono*, 2004, *Tsuchiya et al.*, 2004a, *Tsuchiya et al.*, 2004b, *Caracas and Cohen*, 2005] observed in the lowermost part of the mantle, which is separated as the D" layer.

Due to its importance, much attention was paid to the study of $MgSiO_3$ perovskite. Different groups recorded the Raman spectrum and its evolution with pressure for $MgSiO_3$ perovskite [*Williams et al.*, 1992, *Hemley et al.*, 1992, *Chopelas et al.*, 1992, *Liu et al.*, 1994, *Chopelas*, 1996, *Gillet et al.*, 2000, *Serghiou et al.*, 1998]. The investigations of the physical properties and their evolution with pressure and temperature are only beginning for post-perovskite. For example there is only one study of the lattice dynamical properties [*Tsuchiya et al.,* 2005] which calculates the phonon band dispersion and predicts thermodynamic properties from quasiharmonic approximation. It is the purpose of the present study to determine the Raman susceptibility tensors and to compute the powder Raman spectra and its evolution with pressure for both perovskite and post-perovskite MgSiO$_3$.

**Computational methodology**

All the calculations are based on the local density approximation of the density functional theory (DFT) [*Hohenberg and Kohn*, 1964, *Kohn and Sham*, 1965]. We use the ABINIT package [*Gonze et al.*, 2002, 2005] based on planewaves and pseudopotential. Details of the calculations are reported elsewhere [*Caracas and Cohen*, 2005]. The dynamical matrices, Born effective charges and dielectric permittivity tensors, were computed within density-functional perturbation theory (DFPT), using the responses to atomic displacements and homogeneous electric fields [*Baroni et al.*, 1987, *Gonze et al.*, 1992, *Gonze*, 1997, *Gonze and Lee*, 1997, *Baroni et al.*, 2001, *Gonze et al.*, 2005]. The usual precision of DFT is usually a few percent in bond lengths and better then 20 wavelengths for mode frequencies.

Since most of high-pressure experiments performed in diamond-anvil cells report the Raman powder spectra, for comparison we must average the Raman intensity for the parallel and perpendicular components of the signal for all possible orientations of the crystals in the powder. If we consider an incident laser with frequency $\omega_L$, for each phonon mode with frequency ω, at a given temperature *T*, the average is [*Prosandeev et al.*, 2005, *Placzek*, 1934, *Hayes and Loudon*, 1978]:

$$I_{\parallel}^{powder} = C(10G^{(0)} + 4G^{(2)}); I_{\perp}^{powder} = C(10G^{(0)} + 4G^{(2)}); I_{tot}^{powder} = I_{\parallel}^{powder} + I_{\perp}^{powder}$$

where the pre-factor $C \sim (\omega_L - \omega)^4 [1+n(\omega)]/30\omega$ and $n(\omega)$ is the Boson factor $n(\omega) = 1/[\exp(\hbar\omega/k_B T)-1]$, and $G^{(0)}$, $G^{(1)}$ and $G^{(2)}$ are the rotation invariants:

$G^{(0)} = (\alpha_{xx} + \alpha_{yy} + \alpha_{zz})^2/3$; $G^{(1)} = [(\alpha_{xy} - \alpha_{yx})^2 + (\alpha_{yz} - \alpha_{zy})^2 + (\alpha_{zx} - \alpha_{xz})^2]/2$;

$G^{(2)} = [(\alpha_{xy} + \alpha_{yx})^2 + (\alpha_{yz} + \alpha_{zy})^2 + (\alpha_{zx} + \alpha_{xz})^2]/2 + [(\alpha_{xx} - \alpha_{xx})^2 + (\alpha_{yy} - \alpha_{yy})^2 + (\alpha_{zz} - \alpha_{zz})^2]/3$

The $\alpha_{ij}$, with *i, j = x, y, z* are the components of the Raman susceptibility tensor [*Umari et al.*, 2001, *Cardona*, 1982], which is the first derivative of the electric polarizability tensor $\chi_{ij}$, with respect to the atomic displacements $r_{\alpha\tau}$ performed according to the phonon eigenvector $u_{\alpha\tau}$ and scaled by the unit cell volume $\Omega$:

$$\alpha_{ij} = \sqrt{\Omega} \sum_{\alpha\tau} \frac{\partial \chi_{ij}}{\partial r_{\alpha\tau}} u_{\alpha\tau} \text{ and } \chi_{ij} = \frac{\varepsilon_{ij}^{\infty} - \delta_{ij}}{4\pi}$$

From these we observe that the Raman spectra in a powdered sample depend on the frequency of the incident light, the frequency of the phonon, the temperature and the Raman susceptibility tensor. The calculation of the last quantity has been performed only for a few systems using finite differences [*Prosandeev et al.*, 2005, *Umari et al.*, 2001, *Baroni and Resta*, 1986, *Putrino and Pasquarello*, 2002, *Umari et al.*, 2003] or the second order derivative of the density matrix [*Lazzeri and Mauri*, 2003]. We determine the Raman susceptibilities tensors from DFPT, using the ABINIT implementation [*Veithen et al.*, 2005], based on [*Lazzeri and Mauri*, 2003, *Dal Corso and Mauri*, 1994].

The precision of these calculations is on the order of 10-15 % for the Raman intensity (*e.g.* for Si and quartz).

**Crystal structure and symmetry analysis**

The perovskite phase of MgSiO$_3$ has Pnma symmetry, with four formula units per unit cell. The perovskite structure is thermodynamically stable up to 110 GPa pressure at 0K [*Oganov and Ono,* 2004, *Tsuchiya et al.,* 2004a, *Caracas and Cohen,* 2005]. The structure may be described as a condensation of unstable phonon modes represented by octahedral rotations in R and M [*Glazer,* 1972, *Darlington,* 2002, *Gonze et al.,* 2005] of a cubic ideal undistorted perovskite structure. The perovskite structure is formed by a three-dimensional network of corner-sharing SiO$_6$ octahedra, with Mg atoms in the inter-octahedral void spaces. The Mg and Si atoms occupy the 4*a* and 4*c* Wyckoff positions and the O atoms occupy the 4*c* and 8*d* Wyckoff positions. According to group theory, the 57+3 phonon modes at Γ decompose as irreducible representations: $7A_g + 5B_{1g} + 7B_{2g} + 5B_{3g} + 8A_u + 10B_{1u} + 8B_{2u} + 10B_{3u}$.

The post-perovskite structure has Cmcm symmetry, space group no. 62, with two formula units per primitive unit cell. The structure may be described as a layered structure, where the SiO$_6$ octahedra share an edge along one direction and a corner along the second direction. Mg-O bonds keep the layers together. The Mg and Si atoms occupy the 4*a* and 4*c* Wyckoff positions and the O atoms occupy the 4c and 8f Wyckoff positions. According to group theory, the 27+3 phonon modes at Γ decompose as $4A_g + 3B_{1g} + 1B_{2g} + 4B_{3g} + 2A_u + 6B_{1u} + 6B_{2u} + 4B_{3u}$.

For both structures, except for the $A_u$ modes, all the other *u* modes are infrared active and all the *g* modes are Raman active. For single crystal Raman, the $A_g$ modes are visible no matter the crystal orientation, whereas the B modes are sensitive to the crystal orientation with respect to the polarization of the light. In Raman powder spectra, all the modes are active and integrations must be performed in order to obtain the intensity of each peak.

**Raman Intensity**

Based on the value of the susceptibility, we build the Raman spectra for several pressures for the two structures, choosing an incident laser with 457.9 nm wavelength, and a temperature of 300 K. The theoretical intensity we obtain slightly varies with the choice of frequency of the incident light. These conditions mimic the conditions of most of the experimental determinations of Raman spectra in diamond-anvil cell.

Figs. 1 and 2 show the computed spectra for the perovskite and post-perovskite structures, respectively. The list of the most intense peaks is presented in table 1 at 120 GPa. Our theoretical spectra for perovskite are in excellent agreement with the experimental spectra. At 0 GPa, experimentally the most intense modes are at 382 and 501 cm$^{-1}$ (Gillet et al, 2000), while in our theoretical calculations they lie respectively at 389 and 516 cm$^{-1}$. Experiments performed at higher pressures and temperatures [*Chopelas*, 1996, *Serghiou et al.*, 1998] record the merging of several peaks to three main peaks, consistent with the broadening and overlap of multiple peaks from the theoretical image in Fig. 1. The experimental spectra decrease in intensity with increasing pressure [*Chopelas*, 1996], while in our calculations their intensity increases with pressure.

The post-perovskite structure is characterized by the presence of one strong peak at high frequencies, above 800 cm$^{-1}$. At 60 GPa the most intense peak lies at 854 cm$^{-1}$ and there are five other peaks with relative intensity between 10 and 40 % of the main one. At 90 GPa and above, the Raman is dominated by only one peak, while all the others have intensity below 10 % of the main one (Table 1). The theoretical Raman spectra decrease in intensity with increasing pressure. Up to date there are no experimental measurements to confront our theoretical predictions.

The differences in the number of high-intensity modes present in the spectra of the two structures come from topological differences. In post-perovskite the octahedral breathing mode is Raman active and dominates the Raman spectra because the eigenvector modulates the Si-O bond lengths in phase, thus coupling strongly to changes in dielectric constant to give a high Raman intensity. In simple cubic perovskite this mode is at the Brillouin zone boundary and thus Raman inactive and folded back from the zone boundary in Pbnm perovskite and thus has a small matrix element, so it is weak.

**Summary**

We use the density-functional perturbation theory to compute the Raman spectra with intensities for two high-pressure modifications of MgSiO$_3$ - perovskite and post-perovskite - which are the major constituents of the Earth lower mantle. For perovskite structure, the theoretical spectra are very similar to the experimental ones. Several peaks characterize the spectra with relative intensity above 10% of the most intense one. Due to pressure and temperature these peaks broaden and merge in three sets of larger peaks. For post-perovskite we present a theoretical prediction, due to the lack of experimental work.

The spectra are dominated by one strong peak at high frequencies. The rest of the peaks are visible at 60 GPa, but their intensity decreases below 10% of the most intense peak above this pressure. Thus, Raman spectroscopy can be used successfully to identify the perovskite to post-perovskite phase transition.

## Acknowledgments

These calculations were performed on the Altix computer of the National Center for Supercomputing Applications of the University of Illinois in Urbana-Champaign under grant EAR050012. This research was supported by NSF grant EAR0310139. We use the Bilbao Crystallographic Server for the group theory analysis.

TABLE 1: Theoretical intensity of powder Raman peaks for perovskite (pv) and post-perovskite (ppv) MgSiO3 at 120 GPa.

| pv, P = 120 GPa | | ppv, P = 120 GPa | |
|---|---|---|---|
| cm$^{-1}$ | Int.(%) | cm$^{-1}$ | Int.(%) |
| 423 | 100 | 942 | 100 |
| 578 | 67 | 627 | 8.5 |
| 406 | 27 | 819 | 6.5 |
| 286 | 16 | 597 | 4.8 |
| 258 | 11 | 480 | 4.5 |
| 512 | 7 | 728 | 2.6 |
| 334 | 6 | 651 | 1.6 |
| 475 | 5 | 1054 | 1.4 |
| 444 | 3 | 379 | 0.2 |
| 768 | 1 | 725 | 0.2 |
| 400 | 1 | 471 | 0.1 |
| 498 | 1 | | |
| 654 | 1 | | |

FIG. 1: (Color online) Pressure evolution of the theoretical Raman spectra for MgSiO3 perovskite. We consider ideal powders, with no preferred orientation or shape of the grains. We compute the spectra assuming an incident light with a wavelength of 457.9 nm at 300 K. We convolute with a Lorentzian of 5 cm$^{-1}$ FWHM, approximated for the experimental spectra. The spectra is in excellent agreement with experiment, both in relative intensity and peak positions (Serghiou et al. 2005, Gillet et al. 2000). Note that the shifts in peak positions for theory and experiment are primarily due to inexactness of the theory and not due to thermal effects.

FIG. 2: (Color online) Pressure evolution of the theoretical Raman spectra for MgSiO3 post-perovskite. We consider ideal powders, with no preferred orientation or shape of the grains. We compute the spectra assuming an incident light with a wavelength of 457.9 nm at 300 K. We convolute with a Lorentzian of 5 cm$^{-1}$ FWHM. The spectra are characterized by one strong peak at high situated at high frequencies. There are no experimental measurements at the date of the submission.

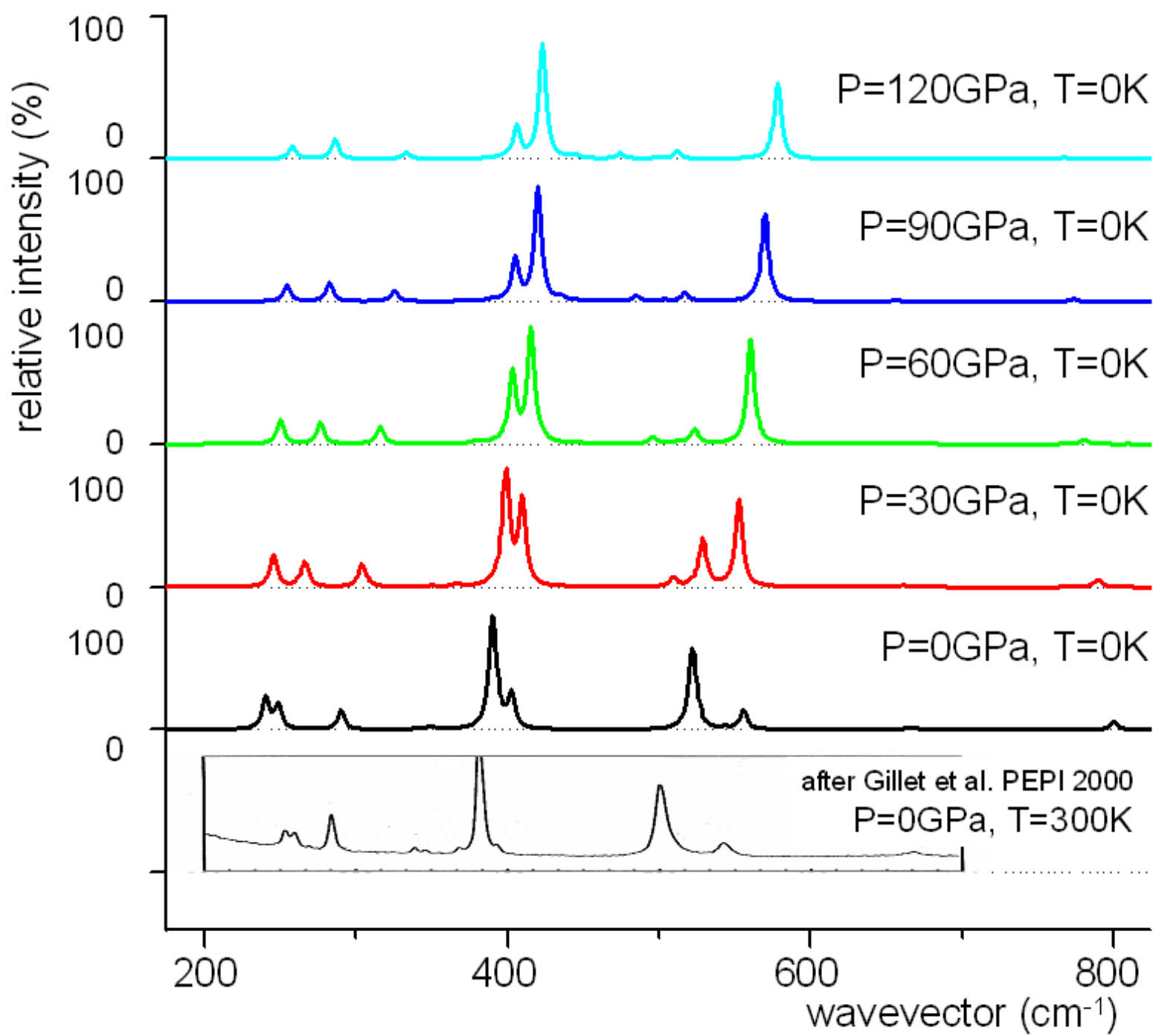

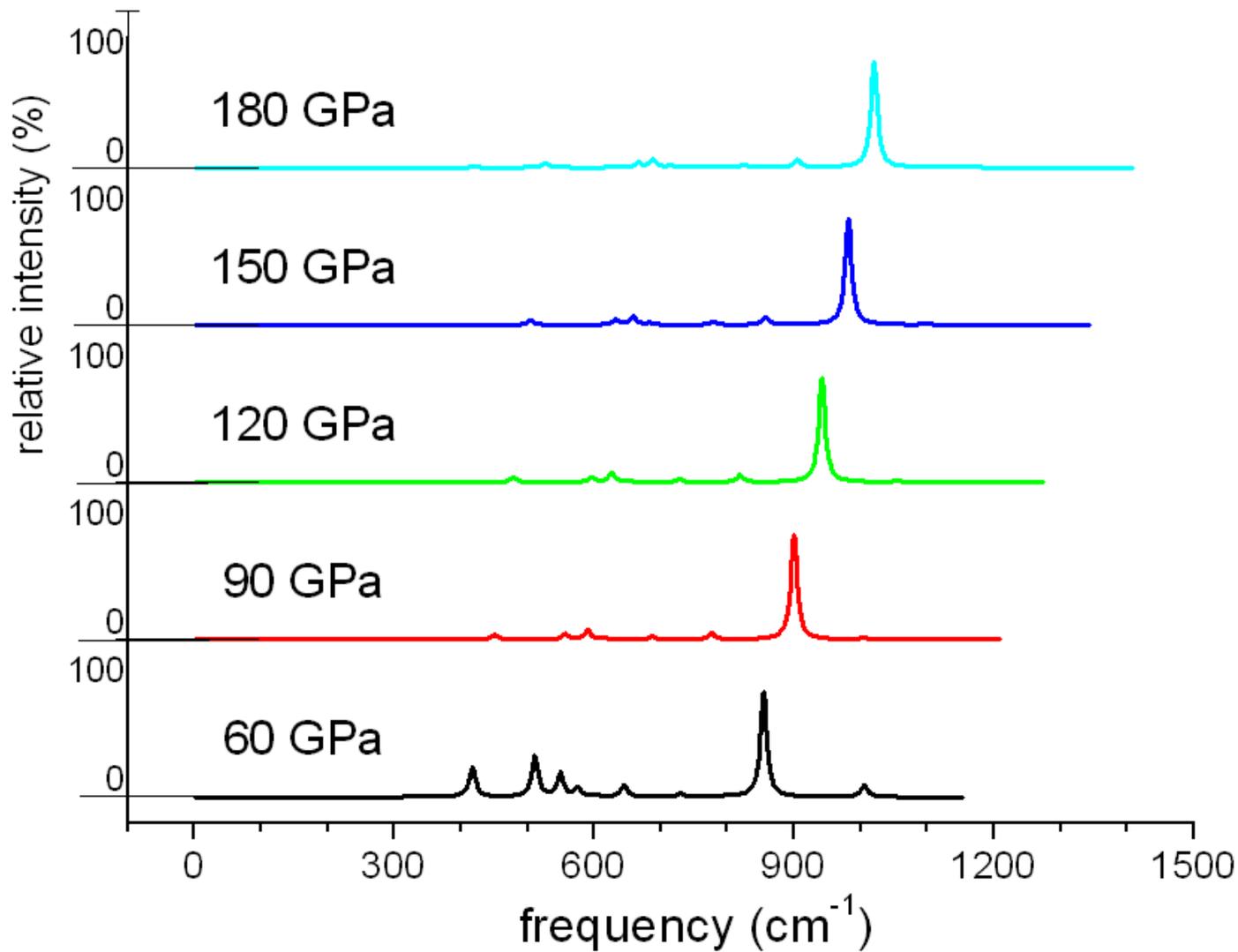